# Transcranial Alternating Current Stimulation (tACS) for patients with Post-Stroke Anomia: Preliminary Data on Picture Naming Performance

Maria Martzoukou[1*], Nefeli K. Dimitriou[2], Binbin Xu[3], Malo Renaud-D'Ambra[4], Anastasia Nousia[5], Anne Beuter[6], Grigorios Nasios[2]

[1] Lab of Cognitive Neuroscience, School of Psychology, Aristotle University of Thessaloniki, Thessaloniki, Greece

[2] Department of Speech and Language Therapy, School of Health Sciences, University of Ioannina, Ioannina, Greece

[3] EuroMov Digital Health in Motion, University Montpellier, IMT Mines Ales, France

[4] Corstim, EuroMov Digital Health in Motion, Montpellier, France

[5] Department of Speech and Language Therapy, University of Peloponnese, Kalamata, Greece

[6] Independent Neuroscience Expert, Montpellier, France

[*] Corresponding author
E-mail: martzoukou@eap.gr

# Abstract

The present study evaluated the effectiveness of transcranial alternating current stimulation (tACS) treating patients with post-stroke anomia using a picture-naming task and a Single-Case Experimental Design (SCED).

A right-handed 38-year-old woman with a left-hemisphere stroke and a left-handed 54-year-old man with a right-hemisphere stroke underwent an eight-week treatment program. Specifically, they participated in a picture-naming task three times a week, alternating between sessions with and without tACS stimulation every two weeks. Electroencephalography (EEG) measurements were taken at the end of each two-week period, and behavioral data were collected before, during and after the treatment. EEG and behavioral assessments were also conducted at one- and three-month follow-ups.

Picture-naming performance was significantly faster during tACS sessions compared to sessions without tACS. By the end of the intervention, both participants demonstrated improved accuracy and speed, with positive effects also observed in behavioral measures. EEG analysis showed that post-treatment brain activity resembled that of healthy individuals performing similar tasks. Patients' improvements in picture-naming and behavioral tests showed that the positive effects remained stable even after three months.

Thus, preliminary data suggest that tACS might be a promising intervention for anomia, with lasting effects. Large-scale studies are needed to confirm these findings.

# Introduction

Aphasia is an acquired language disorder, usually caused by stroke, which negatively affects individuals' daily communication and quality of life [1]. Global stroke events have increased over the last 30 years [2], while stroke incidence rates have risen at younger ages (before 65 years old), leaving working age individuals unable to successfully return to their post stroke jobs and lives [3]. This is partly due to individuals' persistent chronic language impairments that need to be treated [4].

Anomia, a common post-stroke impairment, refers to the difficulty in naming objects, actions, or pictures. Semantic processing, lemma retrieval, and phonological encoding are the core steps in naming and the ones which can be impaired in anomia [5]. Speech therapy is valuable in treating language impairments during initial recovery but plateaus in chronic patients. Therefore, there is a growing interest in noninvasive brain stimulation techniques, such as transcranial direct current stimulation (tDCS) and transcranial alternating current stimulation (tACS).

In tDCS, a low-intensity direct current flows between an anode (positive electrode) and a cathode (negative electrode) on the scalp, modulating neuronal excitability in the brain. Although tDCS shows positive effects for motor function and cognitive rehabilitation after stroke [6, 7], its impact on language recovery remains unclear [8-10].

tACS represents a newer approach in non-invasive neuromodulation technique by applying sinusoidal periodic alternating weak currents that mimic the natural rhythms of brain activity. The adjustment of the alternating current to entrain with the specific brain oscillations allows for a personalized intervention. Entrainment and plasticity are two mechanisms proposed to explain the effects induced by transcranial stimulation [11]. Through entrainment of endogenous activity, tACS could promote rewiring of the brain, soliciting residual plasticity and creating new connections in perilesional regions [12] (see also [13]). Few studies have explored the effect of tACS in patients with stroke, but positive outcomes have been observed in motor [14, 15] and neurological functioning [16]. As for language restoration, there are only three studies investigating the effects of the tACS intervention, one showed that tACS enhances speech comprehension in chronic post-stroke aphasia patients [17], the second reported improved language production abilities that did not, however, reach significance at a group level [18], and the last one reported word-retrieval improvement, but only in a specific phase of tACS condition and stimulation and not in all patients [19]. Thus, further research is required.

The aim of the present study was to explore the effect of tACS on anomia caused by stroke. To this end, a picture-naming task was utilized. Specifically, the present study addressed three questions: (a) Do patients with anomia perform more accurately and faster on a picture-naming task during tACS sessions compared to sessions without tACS? (b) Does a tACS intervention produce positive effects after completion? (c) Do the potential positive effects persist at one- and three-month follow-ups?

The experimental design was of the «n-of-1» type [20], and, specifically, a Single-Case Experimental Design (SCED; [21, 22]). SCED helps overcome major methodological biases in clinical research, such as population size, heterogeneity, and the challenge of finding matched control groups, since each patient serves as their own control-subject. Therefore, SCED is particularly suited for evaluating the efficacy of personalized treatment strategies [22].

In our study, personalization was established through EEG. In particular, prior to tACS treatment, EEG measurements were used to investigate each patient's unique cortical spatiotemporal dynamics. By analyzing the distribution of peaks of activity over time and across different frequency ranges, subject-specific cortical dynamics were mapped. Cortical dynamics involve the synchronization and communication between various brain regions [23]. In picture-naming tasks, alternating activity between occipital and frontal regions has been reported, particularly in the delta frequency range (2-4Hz), which is believed to contribute to the synchronization of cortical processes [24]. Consequently, the objective of the tACS was to enhance cortical dynamics between these regions, improving naming capacity characterized by higher accuracy and faster response times compared to baseline performance.

# Materials and methods

## Participants

Patients had to meet the following criteria for inclusion: a) age between 18 and 70 years old, b) monolingual, native speakers of Greek, c) chronic aphasia following a single cerebrovascular accident, d) difficulty in retrieving names for objects and people, e) motivation and willingness to participate until study completion, and f) a recent brain MRI confirming the stroke lesion site. Exclusion criteria included: a) any contraindication to transcranial electrical stimulation (e.g., intracardiac catheters, cardiac pacemaker, metallic implanted, increased intracranial pressure, pregnancy), b) dementia, c) speech dyspraxia or any other speech motor disorders, d) psychiatric history requiring clinical admission or any behavioral disorders incompatible with brain stimulation, e) history of seizures or epilepsy, f) drug or alcohol dependence, g) visual or hearing deficits, and h) severe comprehension impairment.

Participants' recruitment started on the 1st of October 2022 and ended on the 31st of January 2023. A total of eight patients with aphasia, native speakers of Greek, attending the Clinical Laboratory of Speech and Language Therapy of the University of Ioannina, were screened for eligibility to participate in the study. Evaluations were conducted separately by both a neurologist and an experienced speech-language therapist. Three patients (and their caregivers) declined participation due to time or transportation constraints, while three more were excluded due to comprehension difficulties (n=1) or behavioral disorders (n=2). Ultimately, two patients met the eligibility criteria and consented to participate.

The first participant was a right-handed 38-year-old woman who suffered a left-hemisphere stroke six years prior to the study. The second was a left-handed 54-year-old man who experienced a right-hemisphere stroke nine years earlier. Neurological assessment revealed that the female patient had a typical large infarction of the left middle cerebral artery, with affecting frontal lobe necrosis, while the male patient had an occlusion of frontobasal and precentral branches of his right middle cerebral artery, causing a frontal necrosis, involving putamen and insula. Speech-language evaluation included a medical history review and orofacial / myofunctional assessments. Language abilities were assessed using the Greek Boston Diagnostic Aphasia Examination Shortened Form (BDAE-SF) [25], and naming difficulties were further explored using the Greek Object and Action Test (GOAT) [26] which contains 84 colored photographs (42 actions and 42 objects). The Diadochokinetic Rate assessment was employed to measure the participants' ability to perform rapid and alternating movements of the articulators while repeating specific consonant-vowel sequences (e.g. pa-ta-ka). This assessment aimed to detect any speech sound disorders such as dysarthria or dyspraxia. Lastly, the Edinburgh Handedness Inventory – Short Form [27] was used to evaluate pre-stroke handedness, further confirmed by EGG during naming.

Participants' pre-test performances are presented in Table 1. Both exhibited difficulties retrieving names in "Simple social responses" and "Picture description" subsections of the BDAE-SF (e.g. nerohitis-sink).

**Table 1.** Patients' performance on Boston Diagnostic Aphasia Examination-Short Form (BDAE-SF), Greek Object and Action Test (GOAT)

|  | **BDAE-SF** | **Female** | **Male** |
|---|---|---|---|
| **BDAE-SF** | **II. Auditory comprehension** | | |
| | Word comprehension | 16/16 | 16/16 |
| | Commands | 9/10 | 9/10 |
| | Complex ideational material | 5/6 | 5/6 |
| | **III. Oral expression** | | |
| | Automatized sequences | 4/4 | 3/4 |
| | Repetition of words | 5/5 | 5/5 |
| | Repetition of sentences | 1/2 | 1/2 |
| | Responsive naming | 9/10 | 7/10 |
| | Boston naming test (short form) | 9/15 | 7/15 |
| | Special categories screening | 12/12 | 12/12 |
| | **IV. Reading** | | |
| | Letter and number recognition | 8/8 | 8/8 |
| | Picture-word matching | 4/4 | 4/4 |
| | Word reading | 15/15 | 15/15 |
| | Reading of sentences | 5/5 | 5/5 |
| | Reading of sentences / comprehension | 3/3 | 3/3 |
| | Comprehension of sentences / paragraphs | 3/4 | 4/4 |
| **GOAT** | **Nouns** | 27/42 | 25/42 |
| | **Verbs** | 39/42 | 29/42 |
| **Diadochokinetic Rate** | | No difficulties | No difficulties |
| **Edinburgh Handedness Inventory** | | 87 (right hand) | -75 (left hand) |

All participants and their caregivers were informed about the nature and procedures of the study both verbally and in writing. Written informed consent forms were then signed by both the participants and their caregivers, acknowledging that they could withdraw from the experiment at any time without providing justification. The research protocol was approved by the Ethics Committee of the Medical School of Ioannina, University of Ioannina (approval nr. 49625) and conducted in accordance with the Declaration of Helsinki.

## Materials

Materials were selected from the multilingual picture database by Duñabeitia et al. [28]. Out of 500 Greek words and corresponding-colored pictures (in PNG, 300×300 pixels, 96 DPI), 252 items were selected for this study. Pictures were excluded if they allowed multiple lexical labels (e.g. rabbit or hare), were visually unclear, depicted culturally unfamiliar items (e.g., boomerang), or involved non-Greek-origin words (e.g., drums).

From the 252 items, two were selected as "quality" words to verify data stability: one common and easy-to-pronounce word (miti "nose") and one rare, difficult-to-pronounced word (struθokamilos "ostrich"). To enhance motivation, participants also selected 12 personally meaningful but difficult-to-produce words from the full 500-word database.

Thus, the final set included 264 words, divided into 25 lists of 100 words each. Each list contained 70 different words (presented once), two quality control (each presented 10 times), and one motivational word (presented 10 times). Lists were balanced for difficulty based on an algorithm factoring familiarity, number of syllables, and consonant clusters: (syllables x 50) + (number of consonants x 50) / familiarity. One list was used in "Session 0" for familiarization of the naming task and the procedure, and again one- and three-month follow-ups. The remaining lists were presented in randomized order across sessions.

## Procedure

### *Experimental design*

In the present study, a SCED design was employed, wherein each patient served as their own control-subject [21, 22]. According to the SCED recommendations [21], at least five measurements per phase are required to ensure adequate statistical validity. A longer baseline phase enhances performance stability prior to treatment. Moreover, sequential and randomized treatment introduction across patients is recommended [29].

In alignment with these recommendations, an ABAB design was selected for this study, considering it as a robust design (see Fig 1). The letters A and B correspond to the baseline (A phase) and treatment (B phase) in the ABAB design. In this design, the intervention is re-implemented in a second "B" phase. The ABAB design consists of two main parts: (1) gathering baseline information, applying treatment, and measuring the effects of this treatment; and (2) measuring a return to baseline or observing what happens when the treatment is removed, followed by reapplying the treatment and measuring the subsequent change. This design allows demonstration of experimental control by replicating intervention effects across two treatment phases. Clinicians often prefer the ABAB designs because they conclude with a treatment phase rather than a withdrawal [30].

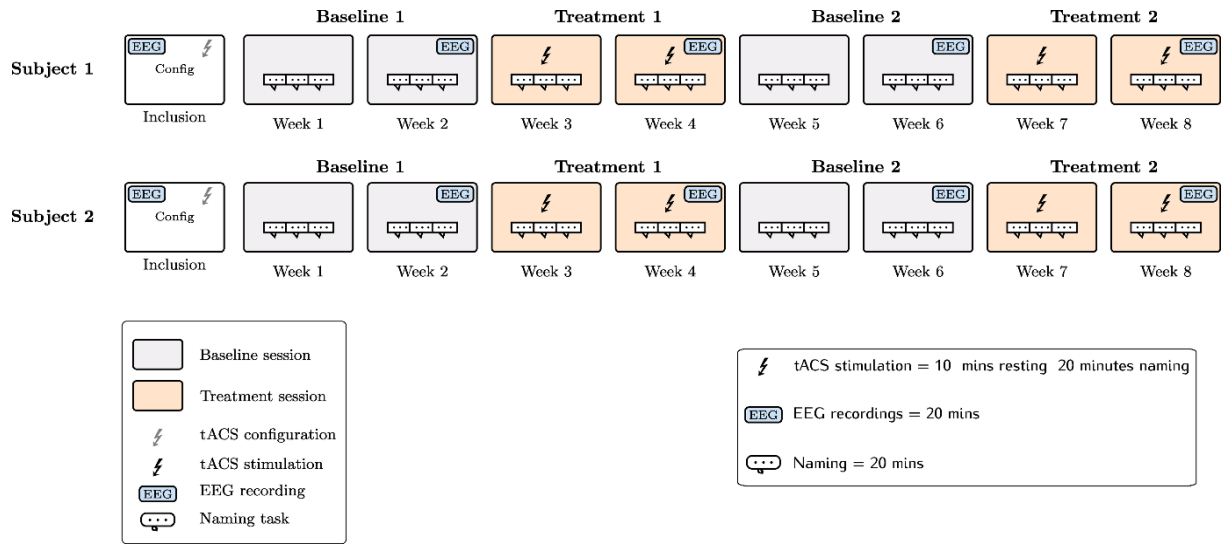

**Fig 1. Schematic of the SCED protocol with repeated alternation of baseline and stimulation phases across sessions.**

Each session includes a fixed sequence of task blocks, with stimulation restricted to shaded intervention blocks and EEG acquired at predefined points to compare baseline, intervention, and washout activity.

In this study, each patient participated in an 8-week protocol comprising 24 sessions. During each baseline phase (A Phase), six repeated measurements (three per week) were collected using naming tasks with 100 words. In the sixth session of each baseline phase, EEG recordings were performed during the naming task. In the treatment phase (B phase), stimulation was applied in three picture-naming sessions during the first week, and in two sessions during the second week, followed by EEG recording in a sixth session. Each tACS session involved 30 minutes of stimulation (10 minutes at rest, 20 minutes during naming). In total, each patient completed 10 simple naming sessions, 4 EEG-combined naming sessions, and 10 tACS-combined naming sessions. The baseline and treatment phases alternated during the study (see Fig 1). Word lists were randomized across sessions. For logistical reasons, the second patient started the protocol one week after the first patient. Post-treatment, naming tasks combined with EEG were administered again, at one- and three-month follow-ups, to assess long-term effects.

Prior to the protocol, participants underwent clinical evaluation, signed consent forms and participated in Session 0, during which stimulation parameters were individually determined based on EEG oscillations. This approach enabled the development of personalized tACS therapy.

*Picture-naming*

Patients sat approximately 60 centimeters from the computer screen, with the experimenter positioned outside their visual field. At the beginning of each trial, a grey screen was displayed for 2 seconds during which participants were allowed to breathe and blink naturally. Patients were then asked to fixate a black cross displayed at the center of the screen for 1.5 seconds while limiting their facial movements. Afterwards, a colored picture was displayed, and patients had to name the object, which was shown for 12 seconds, providing a single response. Oral responses were recorded using an external microphone (see also Fig 2).

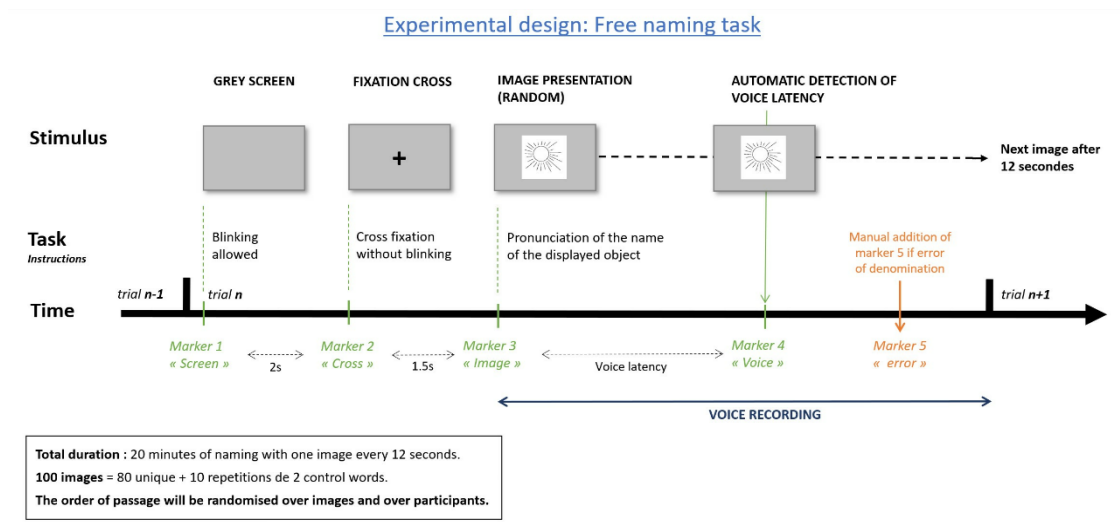

**Fig 2. Experimental design of the free picture-naming task.**
Sequence within each trial: grey screen with free blinking, fixation cross without blinking, random image presentation, overt naming, automatic detection of voice onset, and optional marking of naming errors. Markers 1-5 on the timeline indicate the onset of the grey screen, fixation cross, image, voice response, and experimenter-annotated errors, respectively; one image is presented every 12 s (100 images over ~20 min, in randomized order) while vocal responses are continuously.

The voice onset was automatically detected, and a marker was added accordingly to estimate the naming latency. Each session lasted approximately 20 minutes (100 words x 12 seconds).

*Electroencephalogram recordings*

EEG signals were recorded using the Starstim-32 system (Neuroelectrics®), with precautions taken to minimize electromagnetic and auditory noise. Electrodes were placed on the scalp, and signals were recorded at 500 Hz. Two additional electrooculography electrodes were positioned above and below the right eye to detect eye movements. The reference electrodes (CMS and DRL) were clipped to the right earlobe.

EEG preprocessing followed the pipeline shown in Fig 3, and visual quality inspection was performed. Independent Component Analysis was employed to decompose the EEG signal, facilitating the identification and removal of artifacts such as eye-movement, muscle activity, line noise, and cardiac activity.

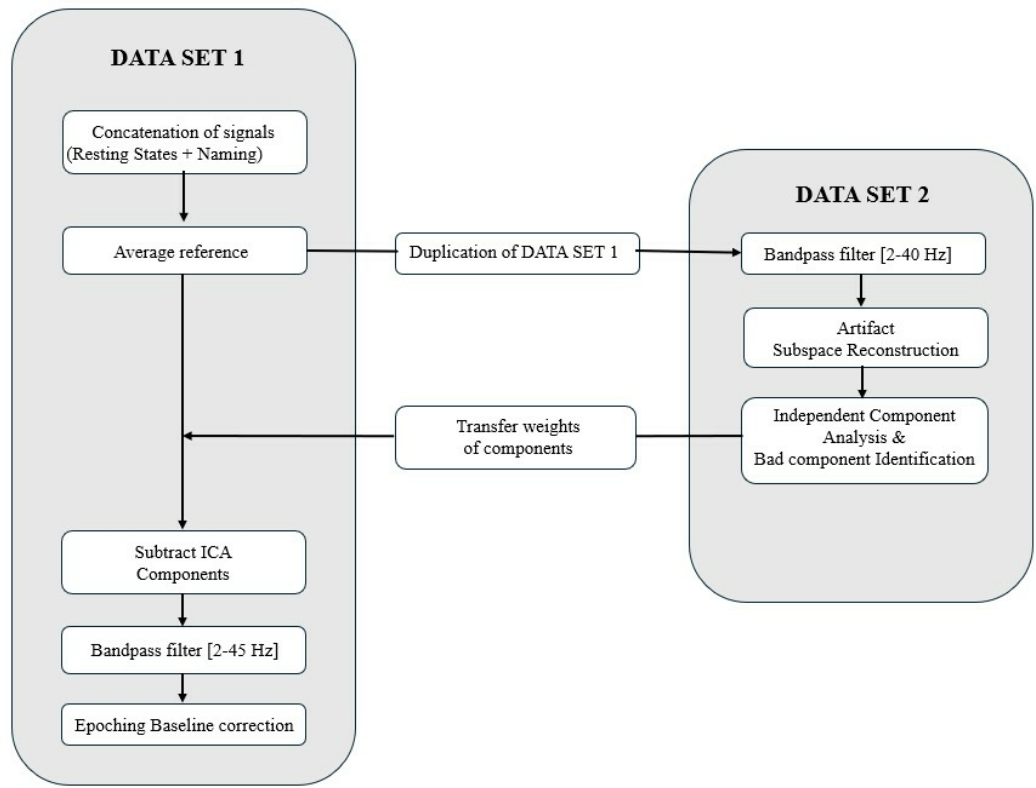

**Fig 3. Preprocessing pipeline for EEG data using two parallel data sets.**
Data set 1 is average-referenced and duplicated into Data set 2 for band-pass filtering (2–40 Hz), Artifact Subspace Reconstruction, ICA and bad-component identification; the resulting component weights are then transferred back to Data set 1 for artifact subtraction, followed by band-pass filtering (2–45 Hz), epoching, and baseline correction.

During Session 0, the stimulation parameters (electrode position, intensity, and frequency) were defined based on each patient's EEG data.

### *Stimulation with tACS*

The tACS stimulation was delivered by a pair of surface circular sponge electrodes (25cm²), soaked in saline solution (0.9% NaCl), and connected to a battery-driven constant current stimulator (Neuroelectrics®, Barcelona, Spain). Electrodes were applied directly to the scalp at the target stimulation areas.

Personalization of the treatment parameters (location, frequency and intensity) was based on EEG results from Session 0. Electrode location was determined by analyzing cortical dynamics using an algorithm developed by Renaud-D'Ambra et al. [24] which extracted the trajectories of the maximum of EEG amplitude over time, illustrating the propagation and main pathways of cortical activity across different frequencies. These trajectories were projected onto a parceled scalp (see Fig 4A) to quantify the likelihood of peak activity in specific cortical regions.

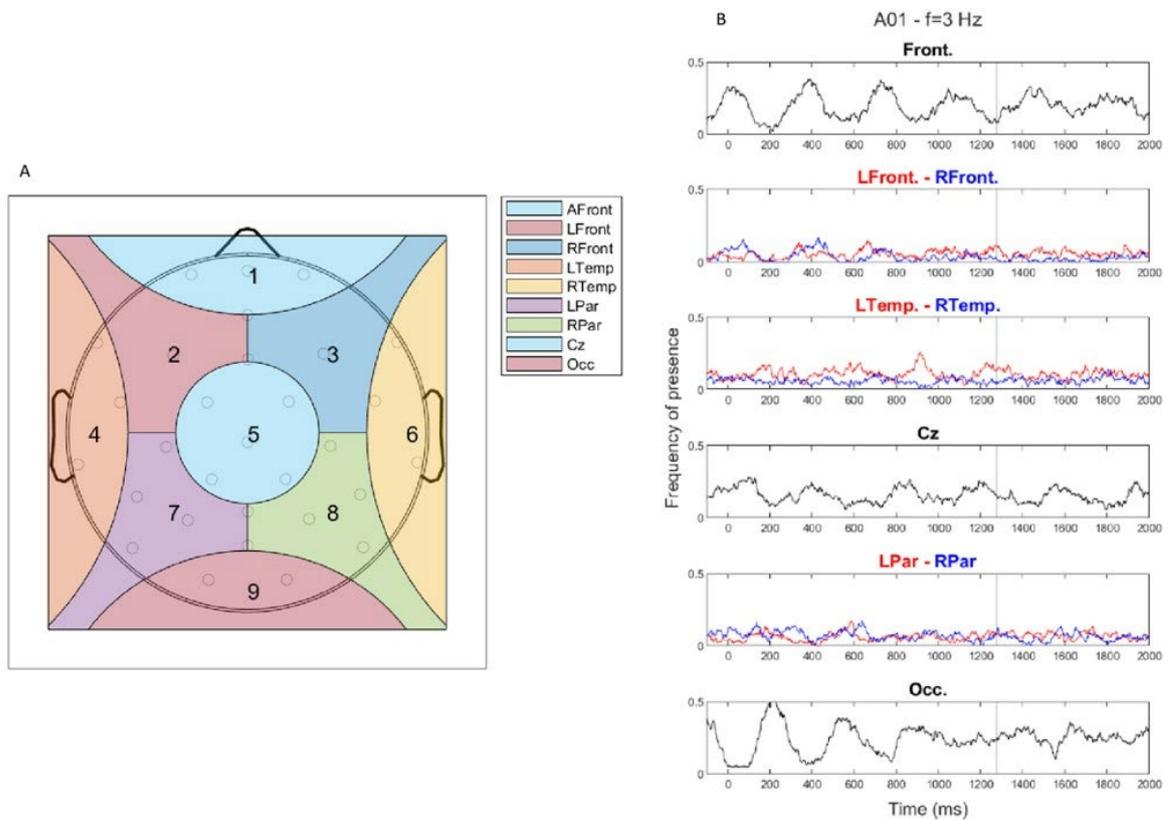

**Fig 1. Electrode placement based on cortical activity trajectories and alternating frontal–occipital amplitude maxima during naming trials.**

A) Scalp parcellation. B) Distribution of the maxima of amplitude across 100 trials at a frequency of 3Hz. 0 indicates the appearance of the image. The vertical line indicates the median latency over all trials. The Y axis indicates the frequency of presence across the 100 trials and the x axis indicates time values in milliseconds.

Fig 4B illustrates the distribution of these amplitude maxima across 100 naming trials for the female participant, revealing a pronounced alternating pattern between the frontal area (FC2) and occipital areas (O1), indicating consistent oscillations of cortical activity between these two main sources. Electrodes were placed accordingly. Using the same procedure, the electrodes placement for the male participant were determined over the left frontal (F1) and right parieto-occipital area (P4) regions (Fig 5).

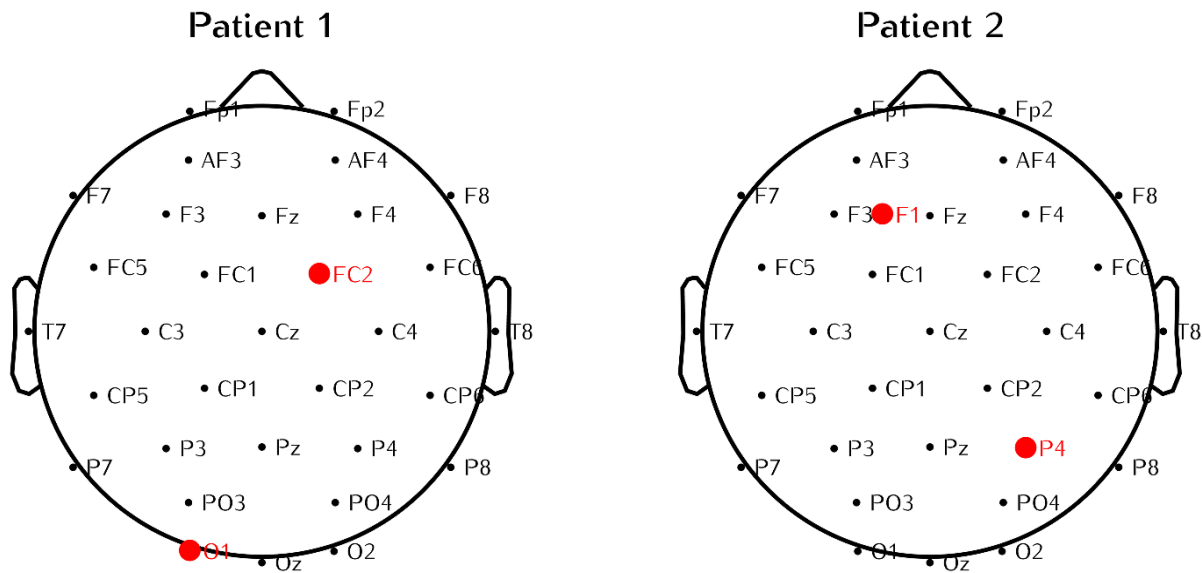

**Fig 5. Stimulation electrode positions for each patient plotted on the 10-20 EEG montage. Red circles indicate the sites selected for stimulation (Patient 1: FC2 and O1; Patient 2: F1 and P4).**

As for frequency, the cortical dynamics between frontal and occipital areas were assessed to identify the salient frequency. This frequency, specific to each patient, was defined as the stimulation frequency. If no alternating pattern between parieto-occipital and frontal areas was detected, based on Session 0, the patient was not retained in the study. For both patients, the frequency was set at 5Hz.

Regarding intensity, the stimulation level was also adaptively determined during Session 0. The initial stimulation for each of the patients was set at 1mA for 5 seconds. If no sensation or discomfort was reported, the intensity was gradually increased by 200μA increments until the appearance of phosphenes (harmless flash of light that may appear during stimulation near the occipital cortex) or mild uncomfortable sensation (i.e., tingling or heat). The intensity was then reduced until a comfortable level was regained. For subsequent stimulation sessions, intensity levels were rechecked to ensure comfort and safety [31]. Final stimulation intensities were set at 1.2 mA for the woman and 2.0 mA for the man.

Data analysis started on July 1, 2024. Only the neurologist of the research team had continuous access to the participants' personal data and full medical records, both during and after data collection.

# Analyses

## Accuracy

For the accuracy and the latency analysis the 70 different words per list were evaluated. In terms of accuracy, participants' responses were recorded and then transcribed by a neuro-linguist trained in transcription. A speech and language therapist conducted a second review to validate the analysis. Discrepancies were discussed and resolved. For error classification, the analysis suggested by Kasselimis et al. [32] was adapted. However, unlike their study, which analyzed short narrations and picture descriptions, our picture-naming task did not expect morphosyntactic errors. Thus, naming errors were classified into three categories: semantic, phonological and neologisms categories. An additional "no answer" category was included to capture "don't know" responses or absence of responses.

Furthermore, for the needs of the present study, three specific error types were added. The semantic error category was expanded by introducing two more types; "hyponym" and "hypernym", describing cases where patients provided a word with a more specific or more general meaning, respectively, compared to the target. Additionally, a "metathesis" type was added under phonological error category to capture cases where sounds within a word were transposed or switched. Dialectical forms were excluded from any further analysis, while in cases in which two phenomena were attested in the same word, both types of errors were annotated. As in Kasselimis et al. [32] though, errors in which more than half of the target word was incorrect, resulting in a non-existing word, as well as cases in which word-responses retained the structure of a Greek word but do not exist were classified under the category "neologisms" (for more information see Table 2).

## Latency

Latencies, defined as the participants' response time from the appearance of the picture until the beginning of their reply, were automatically calculated using Microsoft Azure Cognition service. Instances of unsuccessful starts and hesitations were identified, and the corresponding latencies were excluded from further analysis. Response times were calculated exclusively for correct responses out of the 70 different words per list.

## Statistical analysis

Statistical analysis was conducted to explore whether each patient's performance in terms of accuracy and latency improved during the tACS sessions compared to sessions without tACS. Given that the study followed an 8-week protocol comprising 24 sessions, with additional follow-up assessments at one and three months, and involved repeated measurements rather than isolated time points, performance during tACS and non-tACS sessions was compared using non-parametric independent-samples t-tests. This approach was selected because the data, assessed via the Explore function, were found not to be normally distributed, and the large number of measurements allowed valid application of non-parametric comparisons across conditions. This analysis was conducted on the session-level data for each participant separately. The results of these comparisons are supported by the distribution of the raw data in Fig 6.

## EEG analysis

PsychoToolbox [33], NIC2 software (Neuroelectrics, Barcelona, Spain), EEGLAB [34], SimNIBS [35] and bespoke MATLAB 2021 interfaces were used in the protocol's construction and EEGLab from MATLAB R2023 for the data analysis.

## Patients' behavioral evaluation

Behavioral re-evaluations of the patients were conducted immediately after, as well as one and three months following the completion of the intervention, using the Boston Naming Test and the noun and verb naming subtests of the Greek Object and Action Test (GOAT).

# Results

## Accuracy

The error analysis demonstrated that the two participants adopted different strategies during the naming task. The first patient produced more semantic errors than phonological errors, while the second patient produced more phonological errors than semantic errors. Table 2 presents the percentages (%) of error responses per participant and error type. Interestingly, in one case, the female patient used a plural instead of a singular form, and in seven cases, she used a compound word [/alatopipero/ (salt-and-pepper)] instead of a simple one [/alati/ (salt)]. These errors were classified as morphological errors, accounting for 2.5% of her total errors. For the second participant, eight errors involved attempts to describe the pictures using verbs instead of naming them, constituting 0.8% of his total errors.

Table 2: Percentages (%) of erroneous answers per patient and per type.

| Category | Type | Example | 1st patient (female) | 2nd patient (male) |
|---|---|---|---|---|
| Phonological errors | phoneme substitution | /ma**k**eri/ vs. "ma**x**eri" (knif) | 5.3 | 26 |
| | phoneme omission | /pota/ vs. "po**r**ta" (door) | 1.5 | 24 |
| | phoneme addition | /v**l**ivlio/ vs. "vivlio" (book) | 5.3 | 1.5 |
| | phoneme metathesis | /petaδula/ vs. "petaluδa" (butterfly) | 3.3 | 1.5 |
| | syllable omission/ addition | /ci**t**epelo/ vs. "cipelo" (cup) | 0.3 | 0.5 |
| Semantic errors | similar in form | /anemistiras/ vs. "anaptiras" (lighter) | 6.5 | 2.5 |
| | similar in meaning | /miti/ (nose) vs. "ramfos" (beak) | 17.5 | 8.9 |
| | hyponym | /fuγaro/ (smokestack) vs. "erγostasio" (factory) | 0 | 1 |
| | hypernym | /puli/ (bird) vs. "γlaros" (seagull) | 12.9 | 6.3 |
| | non-similar | /γorγona/ (mermaid) vs. "nerajδa" (fairy) | 16.9 | 5.8 |
| Neologisms | no existing words | /tirigirio/ vs. "plidirio" (washing machine) | 2.5 | 6.2 |
| No answer | | "don't know" or no responses | 25.5 | 15 |

Statistical analyses revealed that both patients' mean performance during the tACS sessions was better compared to their mean performance during the sessions without tACS (Participant 1: 83% with tACS vs. 80% without tACS; Participant 2: 41% with tACS vs. 37% without tACS). These differences, however, did not reach statistical significance for either participant (see also Fig 6).

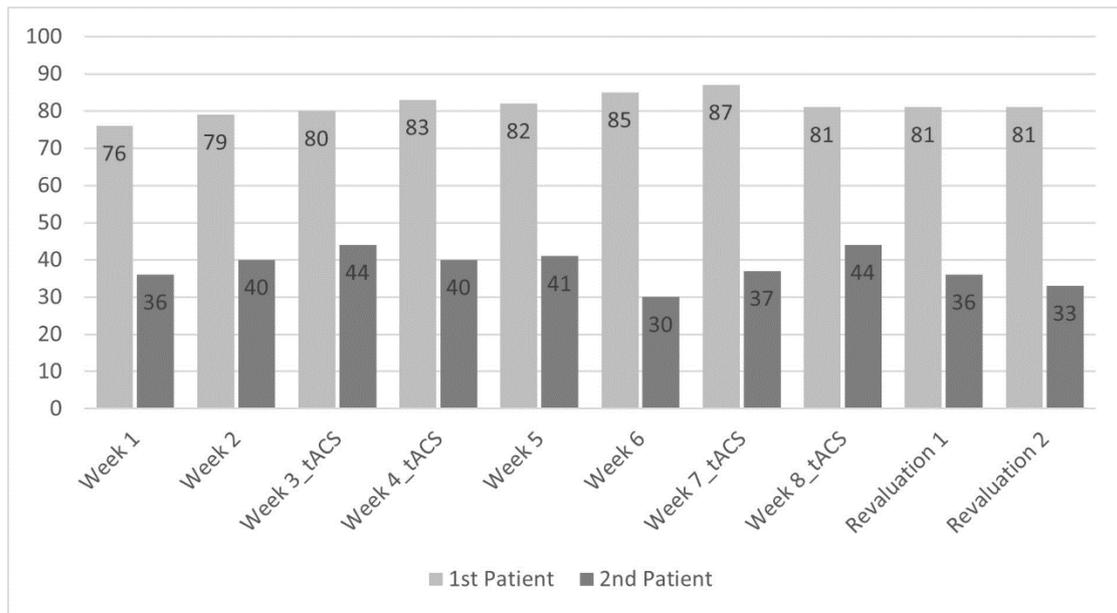

**Fig 6. Weekly percentage performance on the picture-naming task for each patient (Weeks 1–8). Weeks 3, 4, 7, and 8 with tACS stimulation and the first and second follow-up re-evaluations.**

Comparing the first and the last sessions (twenty-fourth) of the intervention revealed that both patients performed more accurately at the end of intervention than at the beginning (Participant 1: 76% → 82%; Participant 2: 31% → 44%). For the female participant, performance one month and three months post-intervention remained improved and stable compared to the initial session (see Fig 6). The male participant's performance one- and three-months post-intervention was similar to his baseline accuracy, with a tendency toward deterioration (see Fig 6).

## Latency

In terms of latency, comparison between the first and the twenty-fourth sessions showed faster responses at the end of the intervention (Participant 1: 684ms → 542ms; Participant 2: 811ms → 674ms). Moreover, the reaction times were faster during tACS sessions compared to sessions without tACS (Participant 1: 556ms with tACS vs. 616ms without tACS; Participant 2: 543ms with tACS vs. 610ms without tACS) (see also Fig 7).

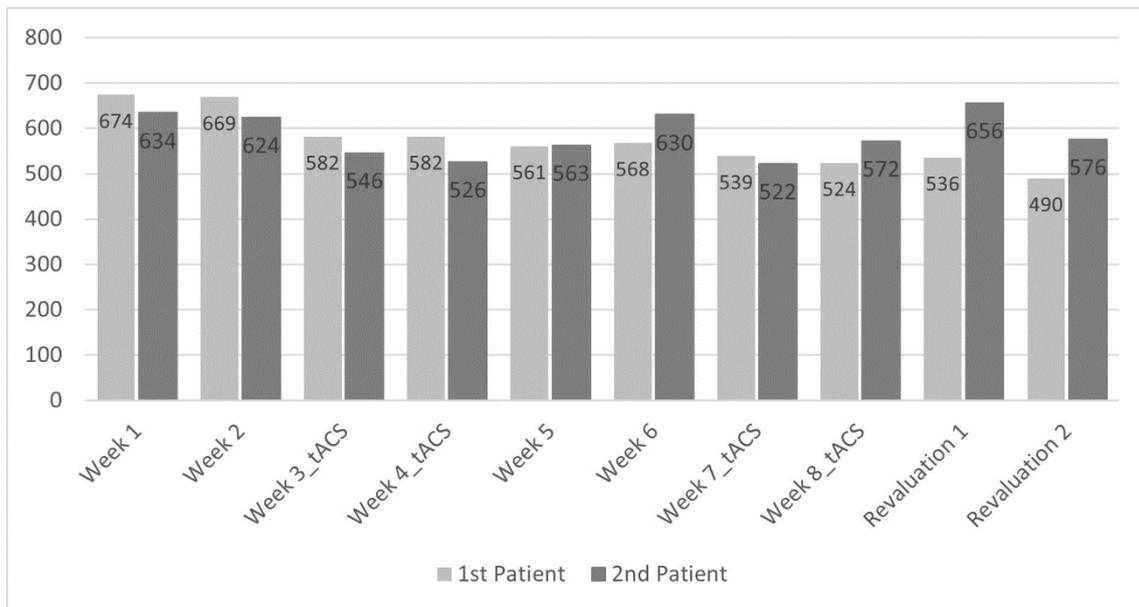

**Fig 7. Mean latencies per week and during the first and the second revaluation for each patient.** Weekly mean naming latencies (ms) on the picture-naming task for each patient (Weeks 1–8), including weeks with tACS stimulation (Weeks 3, 4, 7, and 8). Bars for the first and second follow-up re-evaluations show naming latencies after the intervention period.

This difference reached significance for both patients (Participant 1: Z= 6.040, p <0.001; Participant 2: Z=3.325, p = 0.001). The revaluation at one- and three-months post-intervention showed a tendency for faster latencies for the first participant, while the second participant maintained stable reaction times (see also Fig 7).

## EEG Results

The initial EEG session conducted at the end of the first week (W1 in Fig 8) showed strong desynchronization across the Delta, Alpha, Beta, and Gamma bands. At the end of the fourth week, following two-week intervention with tACS, the EEG revealed more pronounced Delta, Theta, and Alpha activities compared to the initial measurements (W4 in Fig. 8). Similar results were observed at the end of the intervention (W8), and at one- and three-month follow-ups. Throughout nearly all the subsequent sessions, Theta activity remained consistently elevated, and the Alpha activity became stronger compared to the initial session. During the entire study period (except for the initial session), Delta activity was the most dominant, reaching 200% to 300% synchronization compared to the baseline (see Fig 8).

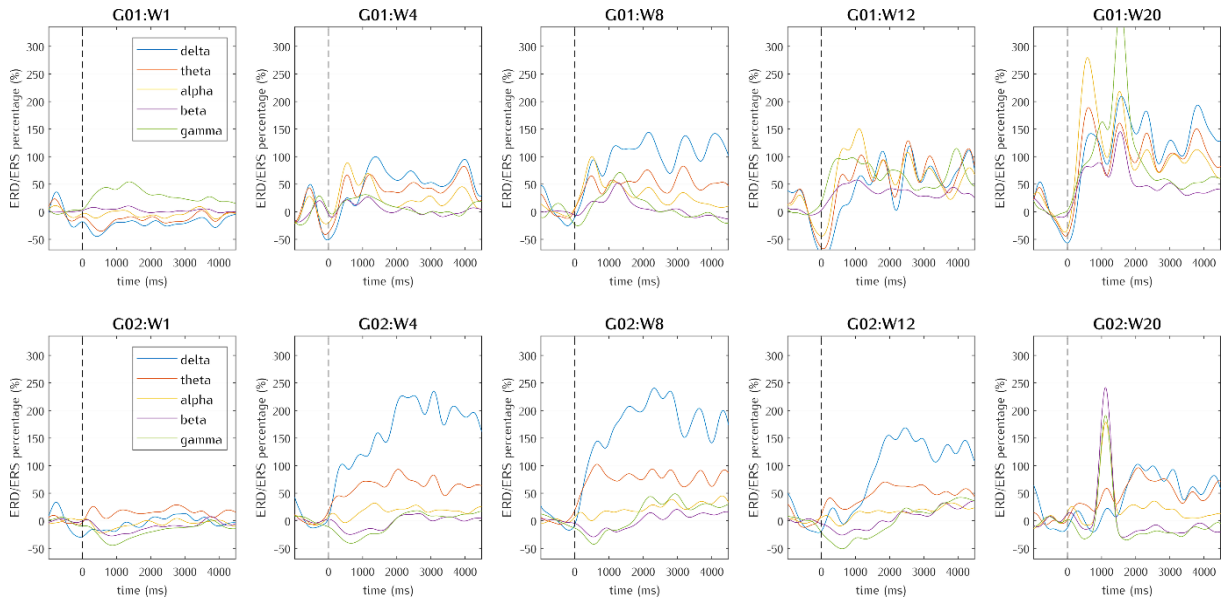

**Fig 8. Time series of median Event-related (De)synchronization.** The end of the first (W1), the median (W4), the last week (W8) and by the two following assessments after one (W12) and three weeks (W20) from the completion of the intervention.

To quantify the spectral changes, Pearson correlations were calculated between the power spectrum of each frequency band across the seven EEG sessions. Higher correlation values indicate greater similarity between the power spectrum profiles. The diagonals (self-correlation) are set to NaN. The most significant spectral changes occurred in the Delta (1-4Hz), Alpha (8-12Hz), and Beta (13-30Hz) bands. The initial session of the treatment showed lower correlations with subsequent sessions, while EEG profiles became increasingly similar over time. High correlations were also observed at one-month (W12) and three-months(W20) follow-ups, especially for the second patient (see Fig 9).

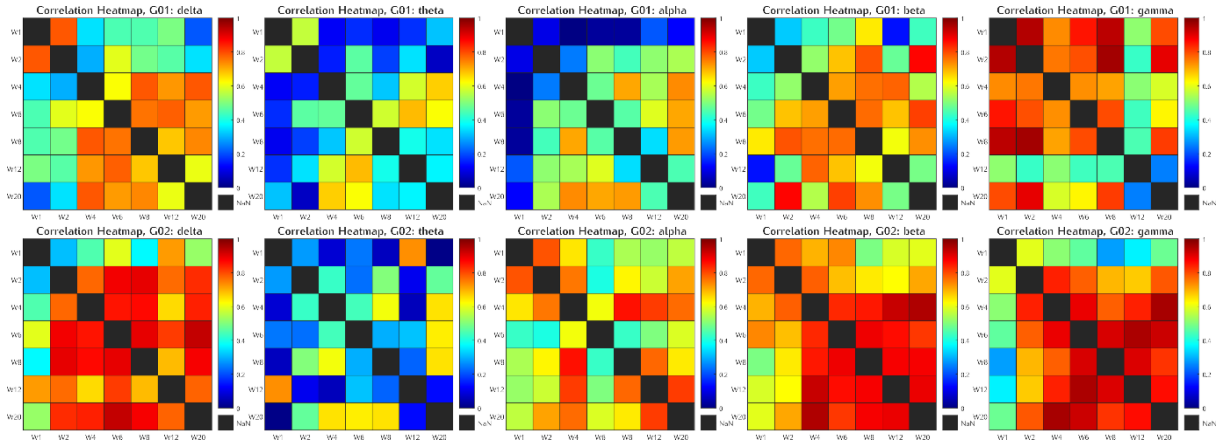

**Fig 9. Pearson correlations between the power spectrum of each frequency band.**
Pearson correlation matrices of band-limited power spectra (delta, theta, alpha, beta, gamma) across seven EEG sessions (W1, W2, W4, W6, W8, W12, W20) for Patient 1 (G01, top row) and Patient 2 (G02, bottom row). Each heatmap displays pairwise correlations (0–1) between sessions for a given frequency band, with diagonal elements (self-correlations) set to NaN.

## Patients' behavioral re-evaluation

Behavioral re-evaluation immediately after the completion of the treatment revealed improved performance for both patients, particularly evident on the Boston Naming Test (short form) subsection of BDAE and the noun and verb naming in the GOAT. For the first patient, performance in the "Simple Social Responses" and "Picture Description" subsections of the first functional section (I. Conversational and Expository Speech) appeared more complete and less effortful compared to the initial evaluation. The second patient also showed improvement in the same subsections. Although difficulties in retrieving names remained evident and disrupted the flow of speech, the patient was more willing to attempt to retrieve the most appropriate word from memory. Ultimately, he was more successful in doing so compared to his initial evaluation. The performance of both patients remained relatively stable at one- and three-month follow-ups (see Table 3).

**Table 3:** Patients' performance on Boston Diagnostic Aphasia Examination (BDAE), Greek Object and Action Test (GOAT) just after, one and three months after the completion of the intervention.

| | | 1st patient (female) | | | 2nd patient (male) | | |
|---|---|---|---|---|---|---|---|
| | BDAE-SF | Just after treatment | 1 month later | 3 months later | Just after treatment | 1 month later | 3 months later |
| BDAE-SF | **II. Auditory comprehension** | | | | | | |
| | Word comprehension | 16/16 | 16/16 | 16/16 | 16/16 | 16/16 | 16/16 |
| | Commands | 10/10 | 10/10 | 10/10 | 9/10 | 9/10 | 9/10 |
| | Complex ideational material | 6/6 | 6/6 | 6/6 | 6/6 | 6/6 | 6/6 |
| | **III. Oral expression** | | | | | | |
| | Automatized sequences | 4/4 | 4/4 | 4/4 | 3/4 | 3/4 | 3/4 |
| | Repetition of words | 5/5 | 5/5 | 5/5 | 5/5 | 5/5 | 5/5 |
| | Repetition of sentences | 2/2 | 2/2 | 2/2 | 1/2 | 1/2 | 1/2 |
| | Responsive naming | 10/10 | 10/10 | 10/10 | 8/10 | 8/10 | 8/10 |
| | Boston naming test (short form) | 15/15 | 15/15 | 15/15 | 12/15 | 13/15 | 12/15 |
| | Special categories screening | 12/12 | 12/12 | 12/12 | 12/12 | 12/12 | 12/12 |
| | **IV. Reading** | | | | | | |
| | Letter and number recognition | 8/8 | 8/8 | 8/8 | 8/8 | 8/8 | 8/8 |
| | Picture-word matching | 4/4 | 4/4 | 4/4 | 4/4 | 4/4 | 4/4 |
| | Word reading | 15/15 | 15/15 | 15/15 | 15/15 | 15/15 | 15/15 |
| | Reading of sentences | 5/5 | 5/5 | 5/5 | 5/5 | 5/5 | 5/5 |
| | Reading of sentences/ comprehension | 3/3 | 3/3 | 3/3 | 3/3 | 3/3 | 3/3 |
| | Comprehension of sentences-paragraphs | 4/4 | 4/4 | 4/4 | 4/4 | 4/4 | 4/4 |
| GOAT | **Nouns** | 40/42 | 40/42 | 40/42 | 31/42 | 31/42 | 30/42 |
| | **Verbs** | 41/42 | 40/42 | 39/42 | 34/42 | 33/42 | 33/42 |

# Discussion

The aim of the present pilot study was to explore the effectiveness of tACS in patients with anomia. To achieve this goal, two patients with anomia following stroke participated in an 8-week treatment period. During this period, they engaged in a picture-naming task three times per week, alternating between sessions without and with tACS stimulation during picture-naming, every two weeks. EEG measurements were conducted at the end of each two-week period, and behavioral data were collected before and after the intervention. The persistence of the intervention's effects was assessed at one- and three-month follow-ups, using both behavioral and EEG measurements.

The results from the picture-naming task indicated that tACS had a positive effect on patients' performance, thus addressing our primary research question. Specifically, patients with anomia were faster and more accurate during tACS sessions compared to sessions without tACS, although the improvements in accuracy did not reach statistical significance. Thus, it could be argued that tACS enhanced brain function, with a stronger effect on reducing patients' reaction times rather than improving their accuracy. These results come in agreement with previous studies conducted in aphasia in which the language production abilities did not improve significantly [17, 18] or improved for specific patients [19], demonstrating again the need for personalized interventions.

The discrepancy, though, between accuracy and latency has been attributed to different factors. Rosato et al. [36], who examined healthy participants, suggested that accuracy did not improve because participants performed at ceiling even before the tACS sessions. In our study, though, participants were individuals with post-stroke aphasia and their baseline performance differed substantially (76% for the female patient and 31% for the male patient), making a ceiling effect unlikely, particularly for the second participant. One possible explanation is that participants' awareness of latency measurements could have led patients to focus more on their speed than on accuracy. On the other hand, it is possible that latency is more sensitive to tACS effects, whereas improvements in accuracy may require more consecutive sessions. Indeed, it has been suggested that increasing the number of tACS sessions, such as one session per day for five days a week, could enhance accuracy apart from latency [37].

Regarding the second research question, both patients appeared to benefit from the personalized tACS intervention. Although no statistical comparisons were made across the full protocol, their performance during the final session was both faster and more accurate compared to the initial session. Similarly, the behavioral data extracted from the BDAE showed improved performance immediately after the intervention.

EEG data supported these results: both patients exhibited stronger brain activity after tACS application, particularly in the Theta band, which remained consistently elevated in all sessions compared to the baseline Theta activity is associated with cognitive processing and memory retrieval demands during tasks like picture naming [38, 39]. Thus, the sustained Theta enhancement might suggest improved language processing abilities. Moreover, the strong Alpha band activities (8-12Hz) observed during tACS sessions are generally considered as a marker of attention and cognitive control [38, 39]. Lastly, the prominent Delta band activity, observed throughout the study except at baseline, may reflect neuroplasticity process that underline neural recovery and reorganization to support language functions [38, 39].

Interestingly, patterns of Delta and Theta activity have been observed in healthy subjects during a forced picture-naming task (participants waited 1.5 seconds before naming the picture) [40, 41]. We observed similarly elevated Delta and Theta activity in both patients following the treatment (see Fig 10). This high similarity may suggest that tACS contributed to the reorganization of brain activity toward a more typical functional pattern seen in healthy individuals. These results, however, should be interpreted with caution, as they are based on only two patients.

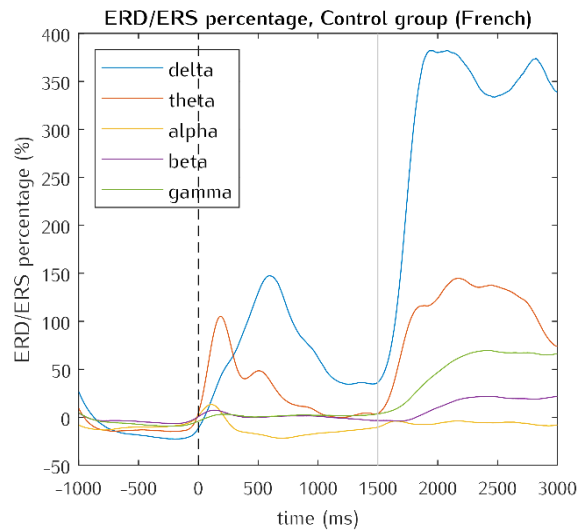

**Fig 10. Average frequency bands found in 16 healthy subjects performing a forced picture naming task.** Event-related desynchronization/synchronization (ERD/ERS) time courses for delta, theta, alpha, beta, and gamma bands in 16 healthy control subjects during the forced picture-naming task. ERD/ERS percentage is plotted relative to baseline and time-locked to picture onset (0 ms, dashed line) and to the go-signal for naming at 1500 ms (grey line).

Concerning the third research question about the persistence of positive effects after treatment, results were mixed. For the first patient (female), accuracy and latency remained improved and stable one and three months after the intervention compared to the baseline. For the second patient (male), accuracy and latency during post-intervention naming sessions were comparable to initial baseline performance, suggesting that the benefits from the tACS intervention were not maintained in picture-name tasks. Consequently, the differences between the two patients in terms of gender, anomia condition, handedness, and stroke lesion location seem to have influenced the duration of the positive effects. This observation suggests that our results cannot be generalized and, thus, more data is needed. Moreover, it highlights the importance of personalized tACS interventions.Behavioral re-evaluations, however, showed positive persistence for both patients. Their performance at one- and three-month reassessments improved compared to baseline and remained stable compared to their performance just after the completion of the intervention. Thus, it is possible that the picture-naming tasks imposed stress on the male patient, whereas, in the BDAE assessment, without strict time constraints, he was able to focus more fully on word retrieval, allowing the positive effects of tACS to manifest in accuracy improvements.

This interpretation is further supported by EEG findings. Post-treatment Delta and Theta activities in these two patients resembled those observed in healthy individuals, and this similarity persisted even three months after stopping tACS. Additionally, higher correlations between EEG frequency profiles across sessions were maintained during one- and three-month follow-ups, particularly for the second patient. This suggests that the neurophysiological effects of tACS may persist beyond the cessation of stimulation, even if behavioral performance under stress-sensitive tasks fluctuates. The overall findings indicate that positive effects of tACS on brain function may not always be fully captured by behavioral tasks sensitive to psychological factors such as stress or anxiety. Future studies should consider assessing behavioral outcomes under varying cognitive load conditions to better disentangle genuine recovery from task-related performance variability.

Even though the intervention with the use of tACs appears to have positive results, mainly in terms of latency and behavioral status, and secondly in terms of accuracy in a picture naming task, it should be taken into account that these outcomes were extracted from only two patients. Therefore, these preliminary findings cannot be generalized, and further replication in a larger sample with more comparable characteristics between patients is needed.

## Conclusions

The overall results of this preliminary study indicate that ten sessions of tACS applied during picture-naming tasks produced positive effects for patients with anomia, with benefits persisting for at least three months post-intervention. In addition to cognitive and language improvements, the intervention appeared to have a positive impact on participants' daily lives. Following the eight-week intervention period without concurrent therapies, the male participant was motivated to resume intensive speech therapy, while the female participant reported greater confidence in her language abilities, resulting in increased sociable engagement.

Thus, while tACS appears to be a promising technique for supporting language recovery after stroke, it must be noted that the intervention was applied to only two participants with differing characteristics, including anomia condition, gender, handedness, and stroke lesion location. Therefore, more research is needed to confirm these findings and to determine the effectiveness of tACS either as a standalone intervention or in combination with other therapies.

**Data availability:** The raw EEG data associated with this study is publicly available :
https://openneuro.org/datasets/ds007315

# Supporting information

**S1 Fig 1. Schematic of the SCED protocol with repeated alternation of baseline and stimulation phases across sessions.**

Each session includes a fixed sequence of task blocks, with stimulation restricted to shaded intervention blocks and EEG acquired at predefined points to compare baseline, intervention, and washout activity.

**S2 Fig 2. Schematic of the SCED protocol with repeated alternation of baseline and stimulation phases across sessions.**

Each session includes a fixed sequence of task blocks, with stimulation restricted to shaded intervention blocks and EEG acquired at predefined points to compare baseline, intervention, and washout activity.

**S3 Fig 3. Preprocessing pipeline for EEG data using two parallel data sets.**

Data set 1 is average-referenced and duplicated into Data set 2 for band-pass filtering (2–40 Hz), Artifact Subspace Reconstruction, ICA and bad-component identification; the resulting component weights are then transferred back to Data set 1 for artifact subtraction, followed by band-pass filtering (2–45 Hz), epoching, and baseline correction.

**S4 Fig 2. Electrode placement based on cortical activity trajectories and alternating frontal–occipital amplitude maxima during naming trials.**

A) Scalp parcellation. B) Distribution of the maxima of amplitude across 100 trials at a frequency of 3Hz. 0 indicates the appearance of the image. The vertical line indicates the median latency over all trials. The Y axis indicates the frequency of presence across the 100 trials and the x axis indicates time values in milliseconds.

**S5 Fig 5. Stimulation electrode positions for each patient plotted on the 10-20 EEG montage.**

Red circles indicate the sites selected for stimulation (Patient 1: FC2 and O1; Patient 2: F1 and P4).

**S6 Fig 6. Weekly percentage performance on the picture-naming task for each patient (Weeks 1–8).**

Weeks 3, 4, 7, and 8 with tACS stimulation and the first and second follow-up re-evaluations.

**S7 Fig 7. Mean latencies per week and during the first and the second revaluation for each patient.**

Weekly mean naming latencies (ms) on the picture-naming task for each patient (Weeks 1–8), including weeks with tACS stimulation (Weeks 3, 4, 7, and 8). Bars for the first and second follow-up re-evaluations show naming latencies after the intervention period.

**S8 Fig 8. Time series of median Event-related (De)synchronization.**

The end of the first (W1), the median (W4), the last week of the intervention (W8) and at follow-up assessments one (W12) and three months (W20) after completion of the intervention

**S9 Fig 9. Pearson correlations between the power spectrum of each frequency band.**

Pearson correlation matrices of band-limited power spectra (delta, theta, alpha, beta, gamma) across seven EEG sessions (W1, W2, W4, W6, W8, W12, W20) for Patient 1 (G01, top row) and Patient 2 (G02, bottom row). Each heatmap displays pairwise correlations (0–1) between sessions for a given frequency band, with diagonal elements (self-correlations) set to NaN.

**S10 Fig 10. Average frequency bands found in 16 healthy subjects performing a forced picture naming task.**

Event-related desynchronization/synchronization (ERD/ERS) time courses for delta, theta, alpha, beta, and gamma bands in 16 healthy control subjects during the forced picture-naming task. ERD/ERS percentage is plotted relative to baseline and time-locked to picture onset (0 ms, dashed line) and to the go-signal for naming at 1500 ms (grey line).